\documentclass[number,preprint,review,10pt]{elsarticle}

\usepackage{amssymb}

\usepackage[margin=4cm]{geometry}

\begin{document}

\begin{frontmatter}



\title{New Instability Results for High Dimensional Nearest Neighbor Search}


\author[label1]{Chris Giannella}
\address[label1]{Dept. of Computer Science, New Mexico State Univ., Las Cruces NM, cgiannel@acm.org}

\begin{abstract}

Consider a dataset of $n(d)$ points generated 
independently from $\mathbb{R}^d$ according to a common p.d.f. $f_d$ with 
$support(f_d)$ $=$
$[0,1]^d$ and $sup\{f_d([0,1]^d)\}$ growing sub-exponentially in $d$.  We prove that: (i) if $n(d)$
grows sub-exponentially in $d$, then, for any query point $\vec{q}^d \in [0,1]^d$ and any $\epsilon >0$, the 
ratio of the distance between any two dataset points and $\vec{q}^d$ is less that $1 + \epsilon$
with probability $\rightarrow 1$ as $d \rightarrow \infty$; (ii) if 
$n(d)$ $>$ $[4(1+\epsilon)]^d$ for large $d$, then for all $\vec{q}^d \in [0,1]^d$ (except a small
subset) and any $\epsilon > 0$, the distance ratio is less than $1 + \epsilon$
with limiting probability strictly bounded away from one.  Moreover, we provide preliminary results
along the lines of (i) when $f_d$ $=$ $N(\vec{\mu}_d,\Sigma_d)$.

\end{abstract}


\end{frontmatter}


\section{Introduction}
\label{intro}

Nearest neighbor search on high-dimensional data is a difficult (and well-studied) problem, in part,
because many commonly used distance functions can exhibit greatly different
behavior in low versus high-dimensional spaces -- a phenomenon often referred to as the ``curse of dimensionality".
In an effort to rigorously analyze this phenomenon, Beyer {\em et al.} \cite{BGRS:1998} defined a nearest neighbor
query with respect to a reference query point $\vec{q}^d \in \mathbb{R}^d$ as {\em unstable} if all of the points
in the dataset are nearly the same distance from $\vec{q}^d$.  In this event, 
the query can be thought meaningless since there is little reason to return any one point over another (see figure 2 
in \cite{BGRS:1998}).  Beyer {\em et al.} (then later others \cite{FWV:2007}, \cite{P:2000}) established sufficient 
conditions on the data generation distributions and dataset sizes under which the probability of query instability 
approaches one as $d \rightarrow \infty$.  Such 
conditions provide useful insight into how the curse can be mitigated 
or must be tolerated as unavoidable.   
We develop a new set of sufficient conditions which improve upon the current ones --
see sub-sections \ref{sec:contributions} and \ref{sec:related} for a description of our contributions and 
their relationship to the literature.

\subsection{Notations and Definitions}

Given $n(.):\mathbb{N} \rightarrow \mathbb{N}$, we represent a 
$d$-dimensional, size $n(d)$ dataset with i.i.d. random vectors
$\vec{Y}_1$, $\ldots$, $\vec{Y}_{n(d)}$ having common p.d.f. $f_d$.    
Let $support(f_d)$ denote the topological closure of
$\{\vec{y} \in \mathbb{R}^d:f_d(\vec{y}) > 0\}$.
Given posative real number $p$, the distance between a pair of points 
$\vec{z}, \vec{w} \in \mathbb{R}^d$ is defined as: 
$||\vec{z}-\vec{w}||_p$ $=$ 
$\left[\sum_{j=1}^d|z_j-w_j|^p\right]^{1/p}$.
Given $\epsilon > 0$, the probability of a nearest neighbor query 
$\vec{q}^d \in support(f_d)$ being unstable is $P_{d,n(.),\vec{q}^d}$ 
$=$ $Pr\left[\max_{i=1}^{n(d)}\{||\vec{Y}_i - \vec{q}^d||_p\}
\leq (1+\epsilon)\min_{i=1}^{n(d)}\{||\vec{Y}_i - \vec{q}^d||_p\}\right].$

The space of all possible query point sequences
is $\prod_{d=1}^{\infty}support(f_d)$.  We say that data distribution 
sequence $\{f_d:d=1,2,\cdots\}$ and dataset size function $n(.)$ admit
nearest neighbor instability if for any $\epsilon > 0$ and any 
query point sequence $\{\vec{q}^d\}$ $\in$ $\prod_{d=1}^{\infty}support(f_d)$, it
is the case that 
$lim_{d \rightarrow \infty}P_{d,n(.),\vec{q}^d} = 1.$  
We say that $\{f_d\}$ and $n(.)$ strongly fail to admit
nearest neighbor instability if there exists $\zeta < 1$ and a
``large'' $\mathcal{Q}$ $\subseteq$ $\prod_{d=1}^{\infty}support(f_d)$, such that
for any $\epsilon > 0$ and for any 
$\{\vec{q}^d\}$ $\in$ $\mathcal{Q}$, it
is the case that 
$lim_{d \rightarrow \infty}P_{d,n(.),\vec{q}^d} < \zeta$.  
Let $\mathcal{Q}^d$ denote the $d^{th}$ component of $\mathcal{Q}$.
We say that $\mathcal{Q}$ is ``large'' if for any $0 \leq \omega < 1$,  
it is the case that, 
$lim_{d \rightarrow \infty}\frac{\omega^d Volume\left(support(f_d)\right)}{Volume(\mathcal{Q}^d)}$ $=$ $0.$
Note, if $support(f_d)$ $=$ $[0,1]^d$, this last condition is equivalent to
$lim_{d \rightarrow \infty}\frac{Volume([0,\omega]^d)}{Volume(\mathcal{Q}^d)}$ $=$ $0.$

A function $g:\mathbb{N} \rightarrow \mathbb{N}$
is said to grow sub-exponentially if $lim_{d \rightarrow \infty}\frac{log(g(d))}{d}$ $=0$.
A sequence of 
functions, $f_d: \mathbb{R}^d \rightarrow \mathbb{R}$; $d = 1,2, \ldots$,
is said to be bounded above sub-exponentially if, for all $d$, 
$sup\{f_d(\mathbb{R}^d)\} \leq g(d)$.

%
%

\subsection{Our Contributions}
\label{sec:contributions}

For any $\{f_d\}$ bounded above sub-exponentially and $support(f_d)$ $=$ $[0,1]^d$, we prove the 
following: (i) if $n(.)$ grows sub-exponentially, then nearest 
neighbor instability is admitted; (ii) if $n(d) > [4(1+\epsilon)]^d$ for large $d$, then (with $p \geq 1$) 
instability strongly fails to be admitted.   Moreover, we describe preliminary results toward
establishing sufficient conditions under which $\{N(\vec{\mu}_d,\Sigma_d)\}$ admits instability.  

\subsection{Related Work}
\label{sec:related}

Beyer {\em et al.} 
\cite{BGRS:1998} established sufficient conditions upon $n(.)$ and
$\{f_d\}$ for the admission of nearest neighbor instability.  
They proved\footnote{They considered
any non-negative distance function and did not restrict query points to reside in $support(f_d)$.} that
instability is admitted if $n(.)$ is constant and $\{f_d\}$ satisfies: 
$lim_{d \rightarrow \infty}Var\left[\frac{||\vec{Y}_1 - \vec{q}^d||_p}{E[||\vec{Y}_1-\vec{q}^d||_p]}\right]$
$=0$, for any $\{\vec{q}^d\}$ (the {\em relative variance} goes to zero).  Pestov \cite{P:2000}, 
proved\footnote{He considered any metric distance function.} 
that (Corollary 5.5) instability is admitted (except for a small set of query point sequences) 
if $n(.)$ is sub-exponentially growing and $\{f_d\}$ satisfies three conditions, most notably, $\{f_d\}$ forms 
a {\em normal Levy family} as defined with respect to the ``concentration of measure'' phenomena. 
Francois {\em et al.} \cite{FWV:2007} proved that instability is admitted (with $\{\vec{q}^d\}$ $=$ $\{\vec{0}\}$) 
if $n(.)$ is constant and each distribution in $\{f_d\}$ has i.i.d. attributes with mean and variance 
not dependent on $d$.  

Our contributions
significantly advance the above results as follows.  Our sufficient conditions allow $n(.)$ to grow
with $d$ (unlike Beyer {\em et al.} and Francois {\em et al.}), are quite broad 
(unlike Francois {\em et al.} who require the data distributions to have i.i.d. attributes), and 
are easy to interpret (unlike Beyer {\em et al.} or Pestov {\em et al.} which leave open 
the question of which data distribution sequences satisfy the relative variance condition or normal Levy 
condition, respectively).  Moreover, we provide results showing that the sub-exponential growth
assumption on $n(.)$ is strongly necessary: if $n(.)$ grows exponentially, then instability fails to be admitted
for a large space of query point sequences.    Finally, we provide preliminary results toward
establishing sufficient conditions for $\{N(\vec{\mu}_d,\Sigma_d)\}$.  To our knowledge, the 
sufficient conditions for this distribution sequence remain unknown. 

Aggarwal {\em et al.} \cite{AHK:2001} considered distance functions with $p$ a positive integer and 
proved that, for 
constant $n(.) = N$ and data distributions with i.i.d. attributes supported on $(0,1)$, 
$C_p$ $\leq$ 
$lim_{d \rightarrow \infty}\frac{E\left[max_{i=1}^{N}||\vec{Y}_i||_p - min_{i=1}^{N}||\vec{Y}_i||_p\right]}{d^{1/p-1/2}}$
$\leq$ $(N-1)C_p$, with $C_p$ a constant not dependent on $d$.  
They argued that
high-dimensional nearest neighbor behavior is sharply different for each of the following
three types of distance functions: $p=1$, $p=2$, and $p \geq 3$.  However, unlike our contributions, 
they do not provide sufficient conditions on instability and they make the restrictive i.i.d. data 
attribute assumption.   
Hsu and Chen \cite{HC:2009} proved\footnote{They consider any non-negative distance function.} that, for 
constant $n(.)$, the relative variance condition of Beyer is a {\em necessary} as well as a sufficient condition
for instability admission.  They go on to develop a basis for empirically testing whether instability
is exhibited.

Shaft and Ramakrishnan \cite{SR:2006} considered the related problem of analytically quantifying the inherent
limits of nearest-neighbor indexing on high-dimensional data.  They proved that, under conditions related to 
those in Beyer {\em et al.}, the performance of a broad class of index structures approaches that of linear
scan as $d \rightarrow \infty$.  In the stochastic geometry literature, Zanger \cite{Z:2003} studied the 
behavior of a general
class of clustering functions as $d \rightarrow \infty$ and established a connection
to the concentration of measure phenomenon.  A more broadly studied problem in this literature 
is the behavior of nearest neighbor structures as the {\em dataset size} goes to infinity and $d$ remains 
constant.  For example, Penrose \cite{P:1999} considered data generated i.i.d. from a continuous p.d.f. with compact
support (and ``smooth'' boundary) and showed that, as $N \rightarrow \infty$, the distance of any
point to its $k$ nearest neighbor converges, almost surely, to a constant not dependent on $N$.  

A vast literature exists on the development of data structures and algorithms for nearest neighbor search,
for brevity, see the discussion and citations in \cite{HC:2009}.

%

\section{Instability Results}
\label{results}

First we develop a lower-bound on $P_{d,n(.),\vec{q}^d}$ 
making no assumptions on $\{f_d\}$ or $n(.)$.  Define 
$\delta(\epsilon,p)$ $=$ $[(1+\epsilon)^p-1]/[(1+\epsilon)^p+1]$ and let
$\gamma \geq 0$.  If
for all $1 \leq i \leq n(d)$, 
$\left|||\vec{Y}_i - \vec{q}^d||^p_p - \gamma\right| 
\leq \gamma\delta(\epsilon,p)$, then
$\max_{i=1}^{n(d)}\{||\vec{Y}_i - \vec{q}^d||_p^p\}$ $\leq$ 
$\min_{i=1}^{n(d)}\{||\vec{Y}_i - 
\vec{q}^d||_p^p\}\frac{[1+\delta(\epsilon,p)]}{[1-\delta(\epsilon,p)]}$ $=$ 
$\min_{i=1}^{n(d)}\{||\vec{Y}_i - \vec{q}^d||_p^p\}(1+\epsilon)^p$.
Thus, $\max_{i=1}^{n(d)}\{||\vec{Y}_i - \vec{q}^d||_p\}$ $\leq$ 
$\min_{i=1}^{n(d)}\{||\vec{Y}_i - \vec{q}^d||_p\}(1+\epsilon).$  Using
this and the fact that $\vec{Y}_1, \ldots \vec{Y}_{n(d)}$ are i.i.d.,

\begin{eqnarray}
P_{d,n(.),\vec{q}^d} &\geq& Pr\left[\forall i, \left|||\vec{Y}_i - \vec{q}^d||^p_p - \gamma\right|\leq \gamma \delta (\epsilon,p)\right] \nonumber \\
&=& \left(1-Pr\left[\left|||\vec{Y}_1-\vec{q}^d||_p^p - \gamma \right| > \gamma \delta (\epsilon,p)\right]\right)^{n(d)}.\label{ineq2} 
\end{eqnarray}

\noindent Our results are reduced to
upper-bounding the probability that a sum of random variables, $||\vec{Y}_1-\vec{q}^d||_p^p$, deviates significantly 
from a fixed value $\gamma$.  To our knowledge, developing a useful bound in the most general case
is not possible.  To get around this problem, we show how our assumptions on $\{f_d\}$ and $n(.)$
allow the dependences between the components of $(\vec{Y}_1-\vec{q}^d)$ to be broken, and thus, open the door 
to applying standard concentration bounds ({\em e.g.} Hoeffding) on the r.h.s. of (\ref{ineq2}).

Assume $\{f_d\}$ is bounded above sub-exponentially and $support(f_d)$ $=$ $[0,1]^d$.
Let $U_1$, $\ldots$, $U_d$ be i.i.d. and distributed uniformly on $[0,1]$.  Let  
$S$ denote $\{\vec{y} \in [0,1]^d$ $:\left|||\vec{y}-\vec{q}^d||_p^p - \gamma \right|$ $> \gamma \delta (\epsilon,p)\}.$ 
There exists 
sub-exponentially growing function $\beta(.)$ such that, 

\begin{eqnarray*}
Pr\left[\left|||\vec{Y}_1-\vec{q}^d||_p^p - \gamma \right| > \gamma \delta (\epsilon,p)\right] &=& \int_{\vec{y} \in S} f_d(\vec{y}) \partial \vec{y} \\
&\leq& \beta(d)\int_{\vec{y} \in S}\partial \vec{y} \\
&=& \beta(d)Pr\left[\left|\sum_{j=1}^d|U_{j} - q^d_j|^p - \gamma \right| > \gamma \delta (\epsilon,p)\right] \\
&\leq& \beta(d)2exp\left(\frac{-2\delta (\epsilon,p)^2\left[\frac{d}{(p+1)2^p}\right]^2}{d}\right).  
\end{eqnarray*}

\noindent The first equality and inequality follow from the fact that $support(f_d)$ $=$ $[0,1]^d$ and $f_d$
is bounded above sub-exponentially, respectively.  The second inequality follows from Theorem 2 of Hoeffding 
\cite{H:1963}.\footnote{With $\gamma$ $=$ $\sum_{j=1}^dE[|U_j-q^d_j|^p]$, $X_j$ $=$ $|U_j - q^d_j|^p$, and
$t$ $=$ $\left(\delta (\epsilon,p)\sum_{j=1}^dE[|U_{j} - q^d_j|^p]\right)/d$.   Clearly $t > 0$.  Also, since 
$support(U_j) = [0,1]$ and $\vec{q}^d$ $\in$ 
$support(f_d) = [0,1]^d$, then 
$0 \leq |U_{j} - q^d_j|^p \leq 1$.  Finally, $E[|U_j-q^d_j|^p]$ $=$ 
$[(q^d_j)^{p+1} + (1-q^d_j)^{p+1}]/(p+1)$ which, for $0 \leq q^d_j 
\leq 1$, obtains its minimum of $1/(p+1)2^p$ at $q^d_j=1/2$.} 
Plugging this bound into the r.h.s. of
inequality (\ref{ineq2}) yields an expression which goes to one as $d \rightarrow \infty$, due to the 
sub-exponential growth assumptions on $n(.)$ and $\beta(.)$.

\section{Dataset Size Assumption}

Now we relax the assumption that $n(.)$ grows sub-exponentially while still assuming that
$\{f_d\}$ is bounded above sub-exponentially and $support(f_d)$ $=$ $[0,1]^d$.  Suppose that, for large
$d$, $n(d) > [4(1+\epsilon)]^d$.  We further assume that $p \geq 1$.  Our goal in this section
is to show that $\{f_d\}$ and $n(.)$ strongly fail to admit instability.  

Fix $99/100 < \zeta < 1$ and define  
$\mathcal{Q}^d$ as $\{\vec{q}^d \in [0,1]^d: Pr[\max_{i=1}^{n(d)}||\vec{q}^d-\vec{Y}_i||_p - 
(1+\epsilon)\min_{i=1}^{n(d)}||\vec{q}^d-\vec{Y}_i||_p \geq 0] \geq 1-\zeta\}$ and $\mathcal{Q}$ $=$
$\Pi_{d=1}^{\infty}Q^d$.  Clearly, for any $\{\vec{q}^d\}$ $\in$ $\mathcal{Q}$, 
$lim_{d\rightarrow \infty}P_{d,n(.),\vec{q}^d}$
$\leq$ $\zeta$.  Hence, all that remains is to show that $\mathcal{Q}$ is 
large, {\em i.e.} for any $0 \leq \omega < 1$, 
$lim_{d \rightarrow \infty}\frac{Volume([0,\omega]^d)}{Volume(\mathcal{Q}^d)}$ $=$ $0.$

Let $\vec{Y}$ be distributed as $f_d$ and be independent of $\vec{Y}_1$, 
$\ldots,$ $\vec{Y}_{n(d)}$.  Define random variables $D_{min}$ $=$ 
$\min_{i=1}^{n(d)}\{||\vec{Y} - \vec{Y}_i||_p\}$,
$D_{max}$ $=$  $\max_{i=1}^{n(d)}\{||\vec{Y} - \vec{Y}_i||_p\}$.  Such
random variables (or related ones) have received considerable study in the stochastic geometry
literature.  Using one such study \cite{LLC:2008}, we prove, in Appendix \ref{appendix}, the following two inequalities
with $Z$ denoting $D_{max} - (1 + \epsilon)D_{min}$:

\begin{equation}
\label{ineq4}
lim_{d \rightarrow \infty}\frac{E[Z]}{d^{1/p}} \geq \frac{1}{100} \mbox{ and }
Volume(\mathcal{Q}^d) \geq \left[\frac{1}{\zeta \beta(d)} \right]\left[\frac{E[Z]}{d^{1/p}} + \zeta - 1 \right].
\end{equation}

\noindent For any $0 \leq \omega < 1$, inequalities (\ref{ineq4}) as
well as the assumptions that $99/100 < \zeta < 1$ and $\beta(d)$ grows sub-exponentially imply 
that $lim_{d \rightarrow \infty}\frac{Volume([0,\omega]^d)}{Volume(\mathcal{Q}^d)}$ $= 0,$ as needed.

\section{Multi-Variate Gaussian Distributions -- Preliminary Results}


We provide preliminary results concerning instability admission over an important
class of distributions that do not satisfy our assumptions above: $\{N(\vec{\mu}_d,\Sigma_d)\}$.  
The following
simple strategy yields a sufficient condition in the case that: $\vec{q}^d=0$, $\vec{\mu}_d = 0$, $p=2$, 
and the number of eigenvalues of $\Sigma_d$ which do not go to zero grows faster than $n(.)$.
Using the eigenvalue
decomposition of $\Sigma_d$, it can be shown that 

\begin{eqnarray*}
& &Pr\left[\left|||\vec{Y}_1||_2^2 - E[||\vec{Y}_1||^2_2]\right| > E[||\vec{Y}_1||^2_2]  \delta (\epsilon,2)\right] \\
&=& Pr\left[\left|\sum_{j=1}^dW_j^2 - E\left[\sum_{j=1}^dW_j^2\right]\right| 
> E\left[\sum_{j=1}^dW_j^2\right]  \delta (\epsilon,2)\right], 
\end{eqnarray*}

\noindent where the
$W's$ are independent and distributed as $N(0,\lambda_j^2)$ with $\lambda_j$ the
$j^{th}$ largest eigenvalue of $\Sigma_d$.  Chebyshev's inequality shows that the r.h.s. of
the equation above is bounded above by 

$$\left[\frac{2}{\delta(\epsilon,2)}\right]\left[\frac{\sum_{j=1}^d\lambda_j^4}{\sum_{j=1}^d\lambda_j^4 + 
2\sum_{1 \leq \ell \neq k \leq d}\lambda_{\ell}^2\lambda_{k}^2}\right].$$  

\noindent Plugging
this bound into the r.h.s. of inequality (\ref{ineq2}), with $\gamma$ $=$ $E[||\vec{Y}_1||^2_2]$, our 
assumptions above on $n(.)$ and the
$\lambda's$ imply that $lim_{d \rightarrow \infty}P_{d,n(.),\vec{q}^d} = 1.$  

Extending the above strategy to $\vec{q}^d, \vec{\mu}_d \neq 0$ and larger growth rates for $n(.)$ seems possible 
utilizing more complex properties of weighted, non-central chi-square distributions.  However, extending beyond $p=2$
seems difficult as only the 2-norm is preserved by orthogonal transformations.  Also, extending beyond multi-variate
Gaussian data distributions seems difficult owing to the fact that independence of the $W's$ depends
upon the Gaussian assumption. 

\appendix

\section{Appendix: Some Proofs}
\label{appendix}

First we prove the left inequality in (\ref{ineq4}):
$lim_{d \rightarrow \infty}\frac{E[Z]}{d^{1/p}}$ $\geq$ $\frac{1}{100}$, where $Z$ $=$ 
$D_{max} - (1 + \epsilon)D_{min}$ $=$ $\max_{i=1}^{n(d)}\{||\vec{Y} - \vec{Y}_i||_p\}$ $-$
$(1+\epsilon)\min_{i=1}^{n(d)}\{||\vec{Y} - \vec{Y}_i||_p\}$.

Theorems 1.1 and 1.2 of \cite{LLC:2008} 
produce an upper-bound on $E[D_{min}]$ and a lower-bound on $E[D_{max}]$, 
respectively.  These combine to yield\footnote{$V_{d,p}$ denotes the 
volume of the unit-ball in $\mathbb{R}^d$
with respect to the $p$-norm.  $\Gamma(.)$ denotes the standard
gamma function.} 

\begin{eqnarray*}
\frac{E[Z]}{d^{1/p}} &\geq& \frac{\Gamma(n(d)+1/d)\Gamma(n(d)+1)}{d^{1/p}\Gamma(n(d))\Gamma(n(d)+1+1/d)3^{1/2}2^{1/d}e^{(1/2d)}||f_d||_2^{2/d}V_{d,p}^{1/d}} \\
&-& \frac{2(1+\epsilon)}{d^{1/p}(n(d)+1)^{1/d}V_{d,p}^{1/d}} - o\left(\frac{1+\epsilon}{d^{1/p}(n(d)+1)^{1/d}}\right).
\end{eqnarray*}

\noindent Thus,\footnote{$lim_{d \rightarrow \infty}||f_d||_2^{2/d} \leq 1$ since $support(f_d) = [0,1]^d$ and sequence $\{f_d\}$ is 
bounded above sub-exponentially.  Also, the ratio of the $\Gamma()'s$ approaches one because of the equality 
$\Gamma(z+1)$ $=$ $z\Gamma(z)$ for any $z \in \mathbb{R}$.  Finally, $lim_{d \rightarrow \infty}(n(d)+1)^{1/d}$ 
$\geq 4(1+\epsilon)$ 
since, by assumption, $n(d)$ $>$ $[4(1+\epsilon)]^d$ for large $d$.} 

$$lim_{d \rightarrow \infty}\frac{E[Z]}{d^{1/p}} \geq
lim_{d \rightarrow \infty}\left(\frac{1}{3^{1/2}d^{1/p}V_{d,p}^{1/d}} - \frac{1}{2d^{1/p}V_{d,p}^{1/d}}\right).$$

\noindent From \cite{HH:2008} (using the fact that $p \geq 1$) and Stirling's approximation\footnote{For large
$z$, $\Gamma(z)$ $\approx$ $exp(-z)z^{z-1/2}(2\pi)^{1/2}$.} of $\Gamma(.)$ (6.1.3.7 in \cite{AS:1964}), 
$lim_{d \rightarrow \infty}d^{1/p}V_{d,p}^{1/d}$ $\leq$ $2(ep)^{1/p}$.  Hence, the above limit is bounded
below by $(1/100)$, as desired.

Now we prove the right inequality in (\ref{ineq4}): 
$Volume(\mathcal{Q}^d)$ $\geq$ 
$\left[\frac{1}{\zeta \beta(d)} \right]\left[\frac{E[Z]}{d^{1/p}} + \zeta - 1 \right]$,
where $\mathcal{Q}^d$ $=$ $\{\vec{q}^d \in [0,1]^d:$ $Pr[\max_{i=1}^{n(d)}||\vec{q}^d-\vec{Y}_i||_p - 
(1+\epsilon)\min_{i=1}^{n(d)}||\vec{q}^d-\vec{Y}_i||_p \geq 0]$ $\geq 1-\zeta\}$ and
$99/100 < \zeta < 1$.

Let $f_Z$ and $f_{Z|\vec{Y}}$ denote the p.d.f of $Z$ 
and the conditional
p.d.f of $Z$ given $\vec{Y}$, respectively.  Since $support(f_d)$ $=$ $[0,1]^d$, then 
$support(f_Z)$ $\subseteq$ $[0,d^{1/p}]$, thus, $E[Z]$ $=$ $\int_{z=0}^{d^{1/p}}zf_Z(z)\partial z$ $\leq$ 
$d^{1/p}\int_{z=0}^{d^{1/p}}f_Z(z)\partial z$ $=$ 
$d^{1/p}\int_{z=0}^{d^{1/p}}\int_{\vec{q}^d\in [0,1]^d}f_{Z|\vec{Y}}(z|\vec{q}^d)f_d(\vec{q}^d)\partial \vec{q}^d\partial z$
$=$ \\ $d^{1/p}\int_{\vec{q}^d\in [0,1]^d}\int_{z=0}^{d^{1/p}}f_{Z|\vec{Y}}(z|\vec{q}^d)f_d(\vec{q}^d)\partial z\partial \vec{q}^d.$
Hence, 

\begin{eqnarray*}
\frac{E[Z]}{d^{1/p}} &\leq& \int_{\vec{q}^d \in [0,1]^d}f_d(\vec{q}^d)\left[\int_{z=0}^{d^{1/p}}f_{Z|\vec{Y}}(z|\vec{q}^d) \partial z\right]\partial \vec{q}^d \\
&=& \int_{\vec{q}^d \in [0,1]^d}f_d(\vec{q}^d)Pr\left[\max_{i=1}^{n(d)}\{||\vec{q}^d-\vec{Y}_i||_p\} - (1+\epsilon)\min_{i=1}^{n(d)}\{||\vec{q}^d-\vec{Y}_i ||_p\} \geq 0\right]\partial \vec{q}^d \\ 
&=& \int_{\vec{q}^d \in \mathcal{Q}^d}f_d(\vec{q}^d)Pr\left[\cdots \right]\partial \vec{q}^d + \int_{\vec{q}^d \in ([0,1]^d \setminus \mathcal{Q}^d)}f_d(\vec{q}^d)Pr\left[\cdots \right]\partial \vec{q}^d \\ 
&\leq& \int_{\vec{q}^d \in \mathcal{Q}^d}f_d(\vec{q}^d)\partial \vec{q}^d + (1-\zeta)\int_{\vec{q}^d \in ([0,1]^d \setminus \mathcal{Q}^d)}f_d(\vec{q}^d)\partial \vec{q}^d \\ 
&=& Pr[\vec{Y} \in \mathcal{Q}^d] + (1-\zeta)Pr[\vec{Y} \in ([0,1]^d \setminus \mathcal{Q}^d)] \\
&=& \zeta Pr[\vec{Y} \in \mathcal{Q}^d] + 1 - \zeta \\
&\leq& \zeta \beta(d)Volume(\mathcal{Q}^d) + 1 - \zeta.
\end{eqnarray*}

\noindent The second inequality follows from the definition of $\mathcal{Q}^d$ and the last
inequality follow from the assumption that $f_d$ is bounded above sub-exponentially.  The desired
inequality follows.


\end{document}